\documentclass[prb,twocolumn,showpacs]{revtex4}
\usepackage{graphicx}
\usepackage{psfrag}
\usepackage{amsmath}
\begin{document}

\title{The Random Field Ising Model on Hierarchical Lattices II: Ground State Critical Properties}


\author{Alexandre Rosas}
\author{S\'ergio Coutinho}
\affiliation{Laborat\'orio de F\'{\i}sica Te\'orica e Computacional \protect \\
Universidade Federal de Pernambuco\\
50670-901, Recife, Pernambuco, Brazil.}

\begin{abstract}
The ground state critical properties of the Random Field Ising Model (RFIM) on the diamond hierarchical lattice are investigated via a combining method encompassing real space renormalization group and an exact recurrence procedure. The local magnetization and the nearest neighbors pair correlation function are exactly calculated. The fixed-point joint probability distribution of couplings and local fields are numerically obtained and analyzed, indicating that the critical behavior of the model is governed by the zero temperature disorder fixed point. The critical exponents associated with the order parameter and correlation length are estimated showing an universal behavior regarding the choice of the initial probability distribution for initial local fields being the symmetric continuous Gaussian or discrete delta-bimodal.
\end{abstract}

\maketitle

\section{Introduction}
The universality of the critical behavior of the Random Field Ising Model (RFIM), with finite dimension, has been object of controversy.  While some authors indicate that the critical behavior of the RFIM, in cubic lattices, does not depend on the considered random field probabilities distribution\cite{swift97}, others argue that the RFIM with Gaussian or bimodal distributions belong to different universality classes\cite{hartmann}  -- such as it occurs for the RFIM in infinite dimensions (mean field approximation)\cite{aharony78}. Both the works\cite{swift97, hartmann} calculate exactly the ground state of the RFIM and apply the finite size scaling approach to get the critical exponents.  According to the real space renormalization group analysis proposed  by Bray and Moore\cite{bray85}, which argues that the critical behavior is governed by a disorder fixed point at $T=0$, these results are valid for finite temperatures.  However, very recently, Duxbury and Meinke\cite{duxbury01} have shown that, within the mean field approach, the RFIM presents distinct behaviors between the ground state and the finite temperature case.  If mean field approach conclusion is applicable to finite dimension systems, the results of Swift \emph{et al. }\cite{swift97} and Hartmann and Nowak \cite{hartmann} will not reflect the RFIM behavior for finite temperature.  However, the mean field approach can be pathological and not valid for finite dimension systems if, for instance, the local field fixed point distribution  is Gaussian\cite{duxbury01}.

In this paper, which follows a previous study of the phase diagram and thermodynamic potentials of the RFIM\cite{alex} (hereafter I), we study the ground state of the RFIM, either with the $\pm H_0$ delta-bimodal probability distribution for the random fields or the Gaussian one, in diamond hierarchical lattices. We use a procedure that encompasses the real space renormalization group approach and an exact recursive process allowing us to get the local magnetizations and the nearest neighbors pair correlation functions.  In the next section, we will describe the model and we present the renormalization equations properties.  The fixed point distribution is obtained and analyzed in section~ \ref{sec:fixed-point-distr}, where we comment on the applicability of the ground state results for systems with finite temperatures.  In section~ \ref{sec:critical-exponents}, we calculate the ground state critical exponents for the RFIM using two distinct methods and compare their results.  Finally, in section~ \ref{sec:conclusion}, we summarize our results.  

\section{The Model}
\label{sec:model}
The RFIM Hamiltonian is given by:
\begin{equation}
  \label{eq:hamilton-rfim}
  {\cal H}=-J\sum_{<ij>} \sigma_i \sigma_j - \sum_i H_i \sigma_i,
\end{equation}
where $J$ represents the ferromagnetic coupling between spins located at sites $i$ and $j$ of a diamond hierarchical lattice and $H_i$ is the random magnetic field that acts on the Ising spin  $ \sigma_i = \pm 1$.  As usual, $<ij>$ indicates that the sum must be done only on first neighbors.  The quenched random fields variables can be chosen from a Gaussian or delta-bimodal probability distribution ${\cal P}_{H_{0}}(H_i)$, with zero mean and variance $H_0$. 

\begin{figure}[htbp]
  \begin{center}
    \psfrag{h}{\LARGE $h$}
    \psfrag{h1}{\LARGE $h_1$}
    \psfrag{h2}{\LARGE $h_2$}
    \psfrag{hb}{\LARGE $h_p$}
    \psfrag{hbarra}{\LARGE $\bar{h}$}
    \psfrag{hl}{\LARGE $h'$}
    \psfrag{hlb}{\LARGE $\bar{h}'$}

    \psfrag{s1}{\LARGE $\sigma_1$}
    \psfrag{s2}{\LARGE $\sigma_2$}
    \psfrag{sb}{\LARGE $\sigma_p$}

    \psfrag{m1}{\LARGE $\mu_1$}
    \psfrag{m2}{\LARGE $\mu_2$}

    \psfrag{Kl}{\LARGE $K'$}
    \psfrag{K11}{\LARGE $K_{11}$}
    \psfrag{K12}{\LARGE $K_{12}$}
    \psfrag{K1b}{\LARGE $K_{1p}$}
    \psfrag{K21}{\LARGE $K_{21}$}
    \psfrag{K22}{\LARGE $K_{22}$}
    \psfrag{K2b}{\LARGE $K_{2p}$}

    \psfrag{decoration}{\LARGE decoration}
    \psfrag{dizimation}{\LARGE decimation}
    \resizebox*{8.5cm}{!}{\includegraphics{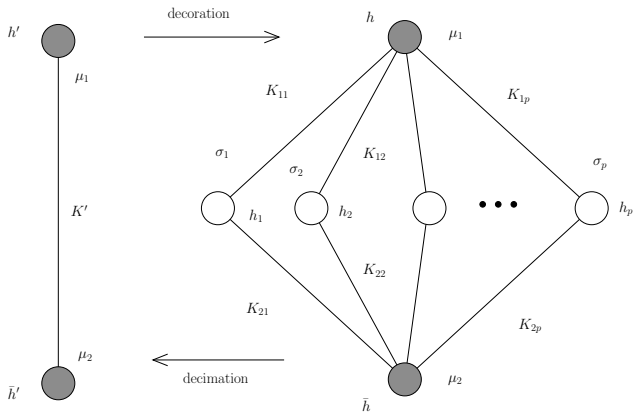}}
    \caption{The diamond hierarchical lattice is built in a recursive process (decoration). Starting from the primitive cell (left), we replace each bond of the preceding hierarchy by the  basic cell (right). The decimation process is done in the spirit of the real space renormalization group in order to obtain the renormalization equations~(\ref{eq:renorJ}) to~(\ref{eq:renorH2}).}
    \label{fig:diam-renor}
  \end{center}
\end{figure}

The diamond hierarchical lattice with $N$ generations, on which the model is defined, is constructed by replacing each bond of the $N-1$ hierarchical lattice by the generating cell, as depicted in figure~\ref{fig:diam-renor}.  The process begins with one bond joining the root sites. This lattice scaling factor equals two and its graph fractal dimension depends on the number of connections ($p$) of the generating cell, that is $d_f = \log (p) / \log(2)$.  For the $p=3$ and $p=4$ cases, considered in this paper, the dimension of the lattice is $2.58 \ldots$ and $3$, respectively.  

Applying the real space renormalization group approach, the $T=0$ coupled renormalization equations for the coupling constants and fields are obtained by decimating the internal sites of the generating cell as sketched in figure~\ref{fig:diam-renor}, giving:  

\begin{subequations}
  \label{eq:renor-geral}
  \begin{align}
  J' & = \frac{1}{2}\sum_{i=1}^{b} \left [ \mathrm{max}(|H_i|,|J_{1i}+J_{2i}|) \right . \nonumber \\
     & \left . - \mathrm{max}(|H_i|,|J_{1i}-J_{2i}|)  \right ] \label{eq:renorJ}\\
  H' & = H + \frac{1}{2}\sum_{i=1}^{b} \left [  \mathrm{max}(|J_{1i}|,|H_i+J_{2i}|) \right . \nonumber \\
     & \left . - \mathrm{max}(|J_{1i}|,|H_i-J_{2i}|) \right ] \label{eq:renorH1}\\
  \bar{H}' & = \bar{H}+\frac{1}{2}\sum_{i=1}^{b} \left [  \mathrm{max}(|J_{2i}|,|H_i+J_{1i}|) \right . \nonumber \\
     & \left . - \mathrm{max}(|J_{2i}|,|H_i-J_{1i}|) \right ].\label{eq:renorH2}
 \end{align}
\end{subequations}

Contrary to the finite temperature case (to cite our work), in $T=0$ the renormalization equations are linear.  In this way, we can divide all the bonds and local fields by a common factor at each renormalization step, without modifying the physical properties of the model.  This property is extremely useful when analyzing the renormalization flow for two main reasons: firstly, it avoids some numerical overflows and secondly choosing the normalization factor as the width of the renormalized field distribution, we keep this distribution with fixed width (unitary) and we get the  fixed point distribution.  This approach was adopted previously by Cao and Machta \cite{cao93} for another three dimensional hierarchical lattice.

The method for evaluation of the local magnetization, which is summarized in I and fully presented in\cite{donato99}, is a generalization of an approach developed to study the local magnetization of the Ising spin glass model on diamond hierarchical lattices under an external magnetic field\cite{donato99}. This approach has also been extensively applied to investigate the local order parameter of several magnetic models defined on hierarchical lattices. Those studies comprise the ferromagnetic Ising model with uniform \cite{morgado90,morgado91} and aperiodic interactions \cite{nogueira2000}, the spin-glass Ising model \cite{nogueira97, nogueira98} and the ferromagnetic Potts model defined on the diamond hierarchical lattices,  as well as the Ising model (both the pure and the spin-glass cases) defined on the Wheatstone bridge hierarchical lattice \cite{camelo99} and on the m-sheet Sierpinskii Gasket fractal lattice \cite{lima99}. 

For the RFIM ground state the coupled recursive equations between local magnetization and pair correlation functions are written as 

\begin{subequations}
  \label{eq:mag}
  \begin{align}
    <\sigma> &= A_1 + A_2 <\mu_1> \nonumber \\
             &+ A_3 <\mu_2> + A_4 <\mu_1\mu_2>\\
    <\sigma \mu_1> &= A_2 + A_1 <\mu_1> \nonumber \\
                   &+ A_4 <\mu_2> + A_3 <\mu_1\mu_2> \\
    <\sigma \mu_2> &= A_3 + A_4 <\mu_1> \nonumber \\
                   &+ A_1 <\mu_2> + A_2 <\mu_1\mu_2>
  \end{align}
\end{subequations}
where,

\begin{subequations}
  \label{eq:contantes}
  \begin{align}
    4A_1&=\mathrm{sgn} (h+K_1+K_2) + \mathrm{sgn} (h+K_1-K_2) \notag \\
    &+ \mathrm{sgn} (h-K_1+K_2) +\mathrm{sgn} (h-K_1-K_2), \\
    4A_2&=\mathrm{sgn} (h+K_1+K_2) + \mathrm{sgn} (h+K_1-K_2) \notag\\
    &- \mathrm{sgn} (h-K_1+K_2) - \mathrm{sgn} (h-K_1-K_2), \\
    4A_3&=\mathrm{sgn} (h+K_1+K_2) - \mathrm{sgn} (h+K_1-K_2) \notag \\
    &+ \mathrm{sgn} (h-K_1+K_2) - \mathrm{sgn} (h-K_1-K_2), \\
    4A_4&=\mathrm{sgn} (h+K_1+K_2) - \mathrm{sgn} (h+K_1-K_2) \notag \\
    &- \mathrm{sgn} (h-K_1+K_2) + \mathrm{sgn} (h-K_1-K_2).
  \end{align}
\end{subequations}
In equations~(\ref{eq:contantes}), $\mathrm{sgn}(x)=1$ if $x>0$, $\mathrm{sgn}(x)=-1$ if $x<0$ and $\mathrm{sgn}(x)=0$ if $x=0$.

The equations~(\ref{eq:mag}) can be used to calculate the whole set of local magnetizations and nearest neighbors pair correlation functions, as a function of the strength of the random fields, by means of an inflation process, as illustrated in figure~\ref{fig:diam-renor}. From these data we can calculate directly certain thermodynamic potentials, such as the average magnetization and the internal energy.  

\section{Fixed Point Distribution}
\label{sec:fixed-point-distr}

Under the renormalization process, the initial distribution of local fields and the single delta function distribution for ferromagnetic couplings evolves either to the ferromagnetic fixed point, if $H_0<H_0^c$, or to the paramagnetic one if $H_0>H_0^c$. In an appropriated distribution parameter space, the flow of the renormalized distributions towards these ``fixed-points'' defines the respective basins of attraction. Within the frontiers of these basins of attraction, one should find the unstable fixed point distribution, which governs the system critical behavior. Therefore, the determination of the fixed point distribution is an important issue to study the critical behavior of the model.  In particular, the form of the fixed point distribution of the random fields determines whether the criticality is actually dictated by the ground state fixed point (as argued by the renormalization group) \cite{bray85} or the result of the mean field approach remains valid in finite dimensions.  

To analyze the renormalization flow, we generate a pool of $10^5$ triplets $ \{ J _ {ij}, H_i, H_j \}$ from the initial distributions, each one corresponding to the coupling constant and the local fields acting of a given bond (see I). We use  equations~(\ref{eq:renor-geral})  to generate a renormalized pool of triplets with same size.  The renormalization sequence of triplets pools determines the renormalization flow.  In the paramagnetic phase ($H_0>H_0^c$), the couplings average between spins vanishes while the width of the local field distribution  ($ \sigma_H = < H^2>^ {1/2}$) grows, but with $<J> / \sigma_H \rightarrow 0$.  On the other hand, in the ferromagnetic phase ($H_0<H_0^c$), the coupling average grows faster than the local fields, so that $<J> / \sigma_H \rightarrow \infty$. In the critical point ($H_0=H_0^c$), however, the transition between the two assintotic behaviors occurs, that is $<J> / \sigma_H$ it is unstablely constant.  Therefore, to determine the fixed point distribution we must find the initial width of the random field probability distribution  for which the renormalized $<J> / \sigma_H$ remains constant.  Using this procedure, we found the critical fields for the three dimensional  lattice  ($H_0^c/J=0.904 \pm 0,001$ for the bimodal distribution and $H_0^c/J=0.997 \pm 0,001$ for the Gaussian one) while, for the lattice with dimension $2,58 \ldots$, we obtain $H_0^c/J=0.587 \pm 0,001$ for the  bimodal distribution and $H_0^c/J=0.610 \pm 0,001$ for the Gaussian one.  These latter values confirm that, at least for the diamond hierarchical lattices, the ferromagnetic long range order is stable for dimensions bellow $d=3$. Furthermore, we have also confirmed that, for $d=2$, there is no phase transition (even an infinitesimal random field distribution width drives the renormalization flow toward the paramagnetic phase).  

\begin{figure}[htbp]
  \begin{center}
    \psfrag{J}{\LARGE $J$}
    \psfrag{H1+H2}{\LARGE $\frac{H_1+H_2}{2}$}
    {\resizebox*{8.5cm}{!}{\includegraphics{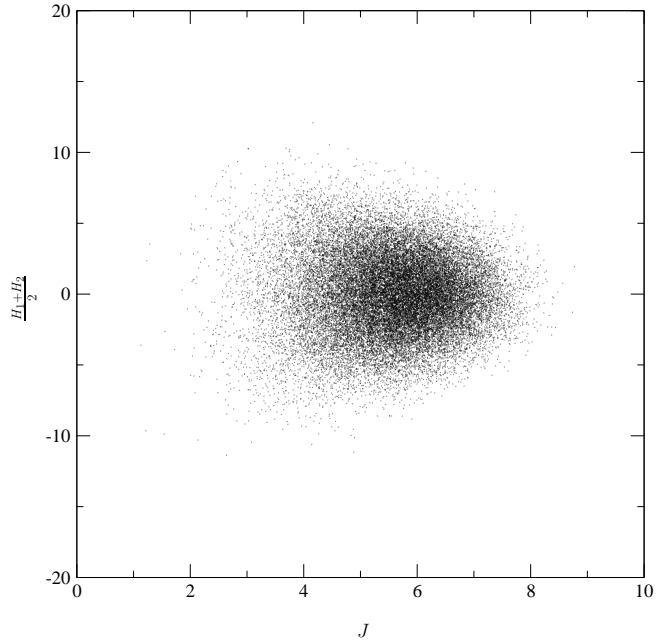}}}
    \caption{Integrated fixed point probability distribution for the delta-bimodal initial random field distribution with width $H_0^c/J =0,903$.}
    \label{fig:dist-crit}
  \end{center}

\end{figure}

\begin{figure}[htbp]
  \begin{center}
    {\resizebox*{8.5cm}{!}{\includegraphics{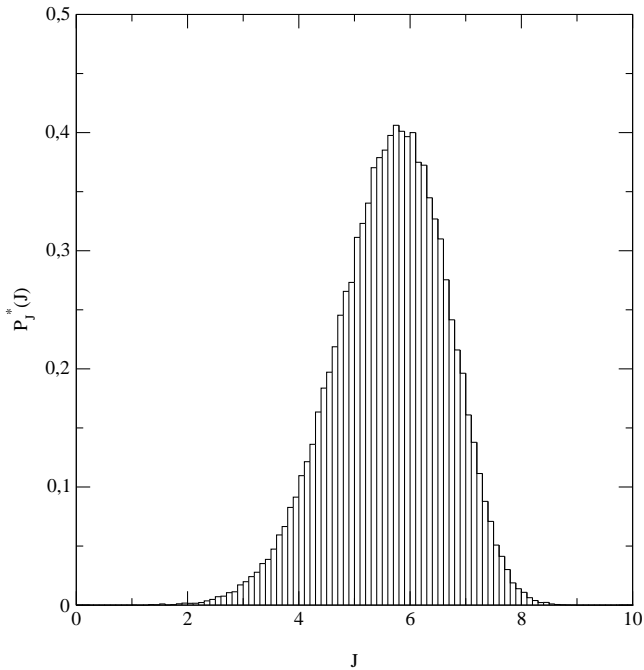}}}
      \caption{Integrated fixed point bond probability distribution for the delta-bimodal initial random field distribution with width $H_0^c/J =0,903$.}
    \label{fig:PJ}
  \end{center}
\end{figure}

\begin{figure}[htbp]
  \begin{center}
    {\resizebox*{8.5cm}{!}{\includegraphics{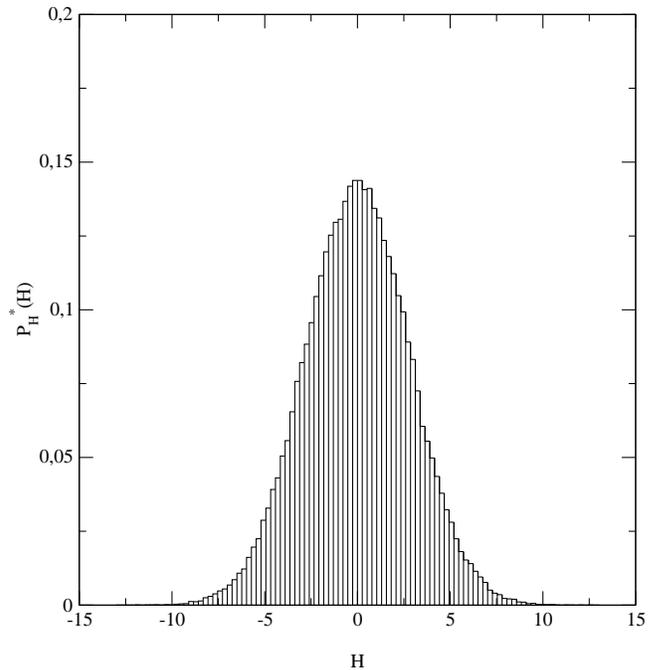}}}
      \caption{Fixed point local field probability distribution for the delta-bimodal initial random field distribution with width $H_0^c/J =0,903$.}
    \label{fig:PH}
  \end{center}
\end{figure}

To generate the fixed point distribution, one may keep the distribution in the fixed point applying, at each step of the renormalization, a small perturbation to the bond distribution, diminishing about 20\% the difference of $<J>$ of the renormalized distribution to its value at the fixed point distribution \cite{cao93}.In figure~\ref{fig:dist-crit} we show the fixed point distribution when the initial random field distribution is the bimodal one. A very similar plot is obtained for the Gaussian initial distribution. It is worth to notice that the fixed point distribution is correlated. This correlation  corroborates the necessity to consider the triplets $ \{ J _ {ij}, H_i, H_j \}$ instead of uncorrelated fields and bonds distributions. In contrast of the result gotten by Cao and Machta \cite{cao93}, the fixed point distribution for the diamond hierarchical lattice, the bond distribution  does not present finite probability in the zero limit of the bond strength, however (see figure~\ref{fig:PJ}) it also presents an asymmetry around $<J>$. Within this aspect, our results are similar to that obtained by Newman \emph{et al. }\cite{newman93}. In the figures~\ref{fig:PJ} and~\ref{fig:PH}, we present the integrated fixed point distribution for the coupling constants and the random fields, respectively, the latter being very close to a Gaussian distribution.  Cao and Machta \cite{cao93} and Newman \emph{et al.  }\cite{newman93} also obtained fields distributions very close to the Gaussian one.  Therefore, despite the bond distribution dependence on the hierarchical lattice, the fixed point field distribution seems to be rather universal.  

To test numerically the Gaussian shape of the (integrated) field distribution, we perform the Lilliefors  test of normality\cite{dudewicz88}. Intuitively, this test measures the `` distance '' between the accumulated probability distribution obtained from the renormalization process and that of the Gaussian. The result of the test shows that the fixed point distribution for the random fields is, at least, very close to a Gaussian, considering a significance level of 0.1\cite{dudewicz88}. Therefore, accordingly to reference\cite{duxbury01}, the results of the mean field approximation are not applicable to the finite dimension system and can we can determine the critical behavior of the RFIM by studying its ground state.  

\section{Critical Exponents}
\label{sec:critical-exponents}

To calculate the critical exponents of the RFIM we use two distinct methods.  The first one is based on the way that the joint probability distribution $ \mathcal{P}(H,J)$ moves away from the fixed point distribution through the renormalization process, while second one is a finite size scaling analysis of the magnetization.  

Following Bray and Moore \cite{bray85}, we define the exponents $x$ (associated with the renormalization of an infinitesimal symmetry breaking field), $y$ (associated with the growth of the width of the distribution) and $z=1 / \nu$ ($ \nu$ being critical exponent associated with  the correlation length).  These three exponents can be directly calculated from the fixed point distribution, as described in the reference \cite{cao93}. Thus, we can calculate 

$$x =  \frac{\log \left (\frac{[\sigma_H']}{[\sigma_H]} \right )}{\log b}, \quad y=\frac{\log \bar{\lambda}}{\log b}$$
and
$$ z = \frac{1}{\nu} = \frac{\log (h'/{h})}{\log b}.$$
where the primes are relate the renormalized quantities, $ [ \ldots ] $ is a configurational average, $b=2$ is the lattice scaling factor, $ \lambda$ is the average of the rescaling factor of the distribution (that is, $ \sigma_H$) and $h = \sigma_H/J-\left (\sigma_H/J \right ) ^*$, $ \left (\sigma_H/J \right ) ^ * $ being the value of $\sigma_H/J$ at the fixed point.  Using these equations, we get the exponents presented in table \ref{tab:xynu}. The exponents $ \alpha$ and $ \beta$ were calculated through the hiperscale relations proposed by Bray and Moore \cite{bray85}.

\begingroup
\squeezetable
\begin{table}[htbp]
  \begin{tabular}{|c|c|c|c|c|}
    \hline
    Exponent & \multicolumn{2}{c|}{Bimodal} & \multicolumn{2}{c|}{Gaussian}\\ 
    &  $d=2.58\ldots$   &  $d=3$    &  $d=2.58\ldots$ &    $d=3$ \\ \hline 
    $x$      &  $2.59 \pm 0.02$   & $2.99\pm 0.02$ & $2.62 \pm 0.04$ & $2.98 \pm 0.02$ \\ \hline
    $y$      &  $1.292 \pm 0.003$ & $1.489\pm 0.002$ & $1.293 \pm 0.003$ & $1.488 \pm 0.002$ \\ \hline
    $\nu$    &  $3.41 \pm 0.06$  & $1.996\pm 0.008$ & $3.41 \pm 0.04$  & $2.00\pm 0.01$ \\ \hline \hline
    $\alpha$ &  $-2.41 \pm 0.09$ & $-1.02 \pm 0.02$ & $-2.41 \pm 0.06$ & $-1.02 \pm 0.02$ \\ \hline
    $\beta$  &  $0.0 \pm 0.2$   & $0.02 \pm 0.04$ & $-0.1 \pm 0.2$   & $0.04 \pm 0.04$ \\ \hline
  \end{tabular}
  \begin{center}
    \caption{The critical exponents $x$, $y$ e $\nu$ were calculated by the analysis or the renormalization flow. $\alpha$ e $\beta$ were calculated with the hiperscale relations proposed in\cite{bray85}.}
    \label{tab:xynu}
  \end{center}
\end{table}
\endgroup

As shown in table \ref{tab:xynu}, all the critical exponents calculated here, either for three dimensional case or to the $d=2,58 \ldots$ one, indicates that the critical exponents do not depend on the choice of the initial probability distribution.  We emphasize that the critical exponent $\beta$ is very small, being possibly zero, for both probability distribution cases.  This almost zero value of $\beta$ may indicate a discontinuity in the magnetization.  The exponent $\alpha$, by it turn, is negative, showing that the specific heat do not to diverge.  These results agree  with the majority of the previous renormalization group results\cite{cao93, newman93, falicov95}. However, our results present an appreciable difference in the value of the exponent $ \nu$ when compared with the value $ \nu=1$ obtained by Boechat and Continentino \cite{boechat90}. The origin of this discrepancy is related with the uncorrelated distribution of bonds and local fields considered by these authors. Therefore, we verify once again the importance of the correlation in the critical properties of the system.  

To calculate the exponents $ \nu$ and $ \beta$ through finite size scaling, we use the equations~(\ref{eq:mag}) to calculate local magnetizations and, then, we calculate the average magnetization 
\begin{equation}
  \label{eq:mag-media}
  M = \frac{1}{N_a} \sum_{\alpha=1}^{N_a} \left | \frac{1}{N_s} \sum_{i=1}^{N_s} <\sigma_i^{(\alpha)}> \right |,
\end{equation}
where $N_a$ is the number of samples and $N_s$ the number of sites of the lattices. The absolute value of the magnetization of a single sample is taken because the random field breaks the symmetry in each sample arbitrarily to the positive or negative magnetization state.  Therefore, the average of the module of the magnetization  is the relevant parameter. 

\begin{figure}[htbp]
  \begin{center}
    \psfrag{M}{\Large $M$}
    \psfrag{H0}{\Large $H_0$}
    \resizebox*{8.5cm}{!}{\includegraphics{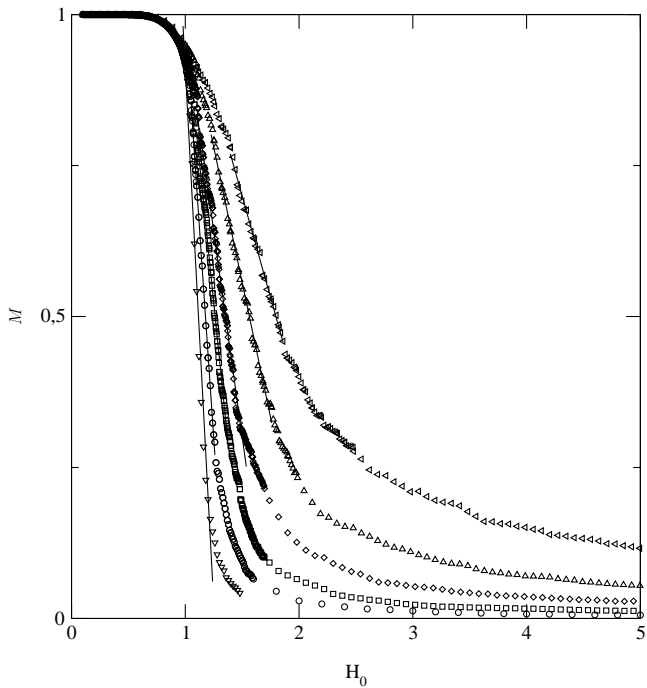}}
    \caption{Ground state magnetization for the RFIM with Gaussian random field probability distribution. The lines are the linear regression used to calculate the critical field for a given lattice size. From above to bellow the curves correspond to lattice with 3 to 8 hierarchies.\label{fig:magfundnorm}}
  \end{center}
\end{figure}

We get the average magnetization for different lattice sizes (3 to 8 hierarchies), as shown in the figure~\ref{fig:magfundnorm}.  Taking a linear regression (represented by the straight lines in the figure~\ref{fig:magfundnorm}), for each lattice size, in the region where the magnetization begins  to fall abruptly until the beginning of the finite size effects, we find the point where this curve intercept the x-axis. We define this point as  the ``critical field '' for that lattice size. Although arbitrary, this criterion is rather robust, since the inclusion or exclusion of some points in limits of the regression region does not modify very much the value of the exponents.  On the other hand, the finite size scaling predicts that all the pseudo-critical fields must converge to the critical field with the same critical exponent $ \nu$\cite{barber83},

\begin{equation}
  \label{eq:nu-scaling}
  \left ( H_0^c(N) - H_0^c \right )/H_0^c \sim N^{-1 / \nu},
\end{equation}
where $N$ it is the lattice size.

\begin{figure}[htbp]
  \begin{center}
    \psfrag{log(N)}{\LARGE $\log(N)$}
    \psfrag{logH(N)}{\LARGE $\log \frac{H_0^c(N) - H_0^c}{H_0^c}$}
    \resizebox*{8.5cm}{!}{\includegraphics{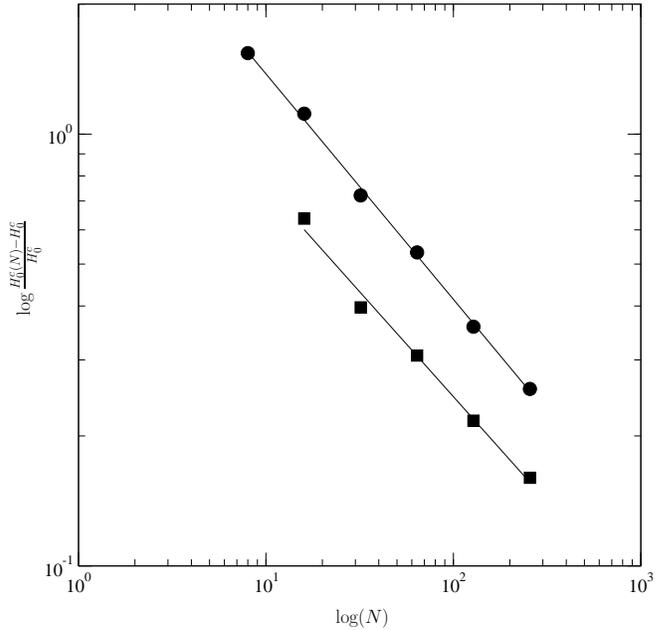}}
    \caption{Illustration of the calculation of the $\nu$ critical exponent. The circles refers to the Gaussian distribution, while the diamonds label the bimodal distribution results.}
    \label{fig:nu-scaling}
  \end{center}
\end{figure}

Therefore, plotting $ \log \left [ \left (H_0^c(N) - H_0^c \right /H_0^c \right ] \times \log (N)$ (figure~\ref{fig:nu-scaling}), we estimate the critical exponent $ \nu$ as  the slope of the best fit of the plot.  With the value of $\nu$ on hands, we obtain the value of $ \beta$ from the scaling law $M \sim N^{-\beta/\nu} f((H_0 - H_0^c)N^{(1/\nu)})$. In the table~\ref{tab:compara}, we present our results for $ \nu$ and $ \beta$ obtained by both considered methods and observe that, within  the error bars, the critical exponents are equal.  

\begin{table}[htbp]
  \begin{center}
    \begin{tabular}{|l|c|c|c|}
      \hline
      \multicolumn{1}{|c|}{Approach} & Exponent & Bimodal & Gaussian \\ \hline
      Renormalization & $\nu$ & $1.996\pm0.008$ & $2.00 \pm 0.01$ \\ \cline{2-4}
      Flow & $\beta$ & $0.02\pm0.04$ & $0.04 \pm 0.04$ \\ \hline
      Finite size & $\nu$ & $2.0 \pm 0.1$ & $1.9 \pm 0.1$ \\ \cline{2-4}
      scaling & $\beta$ & 0.02 & 0.02 \\ \hline
    \end{tabular}
    \caption{Comparison between the critical exponents obtained via the renormalization flow and finite size scaling.}
    \label{tab:compara}
  \end{center}
\end{table}

\section{Conclusion}
\label{sec:conclusion}

In this paper, we study the RFIM on diamond hierarchical lattices, using the real space renormalization group approach.  Studying the renormalization flow, we obtained the joint fixed point probability distribution for coupling and fields.  The analysis of the local field fixed point distribution shows that this distribution is (at least) very close to the Gaussian one.  Therefore, the critical properties of the model are dictated by the disorder critical point (ground state), in contrast to what happens in the mean field approach\cite{duxbury01}.

We calculate the critical exponents of the ground state, showing that the RFIM with delta-bimodal and Gaussian distribution belong to the same  universality class, either for the three dimensional lattice or the  $d=2,58\ldots$ lattice. We found an almost zero $\beta$ critical exponent, indicating a discontinuity in the magnetization.  A scaling analysis with larger lattices could elucidate this question.  However, such analysis exceeds our computational capacities in the present moment.

\acknowledgments

We thank to M. Continentino and J.F. Fontanari for helpful discussions. AR is greatful to CNPq and FACEPE (Brazilian granting agencies) for financial support. This work also received financial support from FINEP (under the grant PRONEX 76.97.1004.00), CNPq and CAPES.

\end{document}